\documentclass[aps,prl,showpacs,twocolumn,groupedaddress]{revtex4}
\usepackage{amsbsy,amssymb,amsmath,bm}
\usepackage{epsfig,wrapfig}
\usepackage{epstopdf}
\input{epsf}

\begin{document}
\title {Josephson phase diffusion in small Josephson junctions: a strongly nonlinear regime}
\author{Mikhail V.~Fistul$^{1,2}$}
\affiliation{$^1$ Theoretische Physik III, Ruhr-University Bochum, D-4081 Bochum Germany}
\affiliation{$^2$ Laboratory of Superconducting Metamaterials and Theoretical Physics and Quantum Technologies Department, National University of Science and Technology MISIS, Moscow 119049, Russia
}

\date{\today}
\begin{abstract}
I present a theoretical study of current-voltage characteristics ($I$-$V$ curves ) of small Josephson junctions. In the limit of a small Josephson coupling energy $E_J \ll k_B T$ the thermal fluctuations result in a stochastic dependence of the Josephson phase $\varphi$ on time, i.e \emph{the Josephson phase diffusion}. These thermal fluctuations destroy the superconducting state, and the low-voltage resistive state is characterized by a nonlinear  $I$-$V$ curve. Such $I$-$V$ curve is determined by the \emph{resonant interaction} of ac Josephson current with the Josephson phase oscillations  excited in the junction. The main frequency of ac Josephson current is $\omega=eV/\hbar$, where $V$ is the voltage drop on the junction. In the phase diffusion regime the Josephson phase oscillations show a broad spectrum of frequencies. The \emph{average} $I$-$V$ curve is determined by the time-dependent correlations of the Josephson phase. By making use of the method of averaging elaborated in Ref. \cite{Gabriele} for Josephson junctions with randomly distributed Abrikosov vortices I will be able to obtain two regimes: a linear regime as the amplitudes of excited phase oscillations are small, and a \emph{strongly nonlinear regime} as both the amplitudes of excited Josephson phase oscillations and the strength of resonant interaction are large. The latter regime can be realized in the case of low dissipation. The crossover between these regimes is analyzed.
\end{abstract}

\maketitle
\section{I. Introduction}
A great attention is devoted to an experimental and theoretical study of small Josephson junctions \cite{Tinkham}. In these systems one can observe such interesting physical phenomena as superconductor-insulator phase transition \cite{Haviland}, Coulomb blockade of Cooper pairs \cite{Aver-Likh,Matveev}, incoherent and coherent Josephson phase-slips \cite{AH,Mooij,Astafiev}, Josephson phase diffusion \cite{MartKautz,FUK}, just to name a few. The physical origin of all these phenomena is the presence of thermal and/or quantum fluctuations that greatly influence the dc and ac Josephson effect. In this paper we consider  moderately small Josephson junctions as the charging energy $E_c$ is smaller than the Josephson coupling energy, $E_J$. For such Josephson junctions one can safety neglect the quantum fluctuations of Josephson phase.  However, as $E_J$ is small, i.e. $E_J \ll k_B T$, the \emph{Josephson phase diffusion regime} induced by thermal fluctuations, occurs.   In the regime of a strong dissipation the Josephson phase diffusion regime has been studied in detail experimentally and theoretically \cite{MartKautz,FUK,Vion-96,IngNaz,GrIngPaul}. Most pronounce features of the Josephson phase diffusion  are the absence of the zero-voltage superconducting state, nonlinear current voltage characteristics ($I$-$V$ curves) occurring in a low voltage region, and a strong suppression of the maximum current value.

In the presence of Josephson phase diffusion dc $I$-$V$ curves can be qualitatively explained as follows. As the dc voltage $V$ is applied the ac Josephson current with the main frequency $\omega=2eV/\hbar$ is flowing in the junction. Such a Josephson current excites the \emph{Josephson phase oscillations } which, in turn, resonate with the alternating part of the Josephson current leading to the finite dc current $I$.
The thermal fluctuations result in a broad spectrum of Josephson phase oscillations and determine the strength of resonant interaction.

It is also  well known for many years that in the Josephson phase diffusion regime the dc $I$-$V$ curves depend crucially on the Josephson phase damping. Such a damping is determined mostly by various dissipative effects and, in particular, the quasi-particles resistance.  In the limit of a large dissipation (damping) the amplitudes of excited Josephson phase oscillations are small, and therefore, using the perturbation analysis the dc $I$-$V$ curve has been carried out quantitatively \cite{Tinkham,FUK,IngNaz}.
\begin{equation}
I=\frac{I_c}{\alpha} \frac{VV_p}{V^2+(\delta V_p)^2},
\label{IVcurve-largedissipation}
\end{equation}
where we introduce the characteristic voltage $V_p=\hbar \omega_p/2e$, the plasma frequency $\omega_p$, and the dimensionless parameter $\alpha$ describing the dissipative effects. In the Josephson phase diffusion regime the thermal fluctuation induce a stochastic part of the Josephson phase $\psi(t)$, and an \emph{average} dc $I$-$V$ curve is determined by the specific time-dependent correlation function of $\psi(t)$, i.e.
$\rho(t)~=~<\cos(\psi(t)-\psi(0))>$. As the damping is large the $\rho(t)$ shows a diffusive form: $\rho(t)~=~\exp(-\delta t)$. The typical $I$-$V$ curve of a small Josephson junction in the Josephson phase diffusion regime is presented in Fig.1.

\begin{figure}[tbp]
\includegraphics[width=2.2in,angle=0]{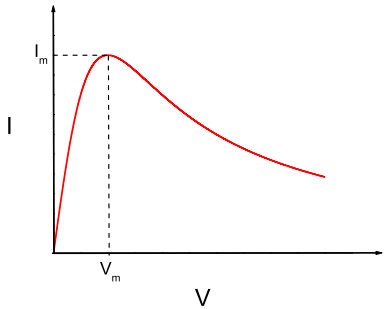}
\caption{The typical $I$-$V$ curve of a small Josephson junction in the Josephson phase regime.
The voltage drop $V_m$ and the current $I_m$ corresponding to the maximum of $I$-$V$ curve are shown.
} \label{Schematic}
\end{figure}

Notice here, that a crucial condition allowing one to obtain Eq. (\ref{IVcurve-largedissipation}) is a large value of the damping parameter, $\alpha \gg 1$. Thus, a next question naturally arises: how vary the $I$-$V$ curves in the limit of a small damping? In such a case the Josephson phase displays oscillations with a large amplitude, and the perturbation analysis can not be applied. Instead of the perturbation approach I will use the method of averaging elaborated in Refs. \cite{Gabriele,FL}. Although this method has been used, previously,  in order to analyze the current resonances in long Josephson junctions with randomly distributed Abrikosov vortices, i.e. coordinate-dependent inhomogeneities, it is possible to adjust such a method to the Josephson junction with thermal fluctuations, i.e time-dependent inhomogeneities.

The paper is organized as follows. In Section II the dynamics of the Josephson phase in the low voltage resistive state and in the presence of thermal fluctuations will be analyzed. In Section III we calculate the time-dependent correlation functions of the Josephson phase $\psi(t)$ determining the electrodynamic properties of small Josephson junction. In Section IV, by making use of the averaging method elaborated in Ref. \cite{Gabriele} we obtain the dc $I$-$V$ curves of small Josephson junctions in the Josephson phase diffusion regime. The Section V provides discussion and conclusions.
\section{II. The dynamics of the Josephson phase in the resistive state: the Josephson phase diffusion regime}

In order to quantitatively analyze the $I$-$V$ curve of a small Josephson junction in the Josephson phase diffusion regime we write the dynamic equation for the Josephson phase $\varphi(t)$
\begin{equation} \label{GenEq}
\ddot \varphi(t)+\alpha \dot \varphi(t)+\sin\varphi(t)= j +\xi(t).
\end{equation}
Here, $j$ is the dc current, and $\xi(t)$ is a random function of time $t$ describing thermal fluctuations (the Langevin force). The dimensionless units were used, i.e., the time is normalized to $\omega_p^{-1}$, the dc bias $j=I/I_c$ is normalized to the
critical current value $I_c$.
The solution of this equation corresponding to the resistive state is written as
\begin{equation} \label{Josphase}
\varphi(t)=vt+\psi(t)+\varphi_1(t)~~,
\end{equation}
where the dc voltage drop $V$ is normalized to $V_p$ as $v=V/V_p$, and the random function $\psi(t)$ determines the Josephson phase diffusion. As the Josephson phase oscillations term $\varphi_1(t)$ is small, the perturbation approach can be used, and the Eq. (\ref{IVcurve-largedissipation}) is recovered. In a generic nonlinear case $\varphi_1(t)$ is written as
\begin{equation} \label{Josphase-general}
\varphi_1(t)=A(t)e^{ivt}+B(t)e^{-ivt},~~ B(t)=A^{\ast}(t)~.
\end{equation}
Thus, $\varphi_1(t)$ shows rapid oscillations of frequency $v$ and a smooth time-dependence describing by the function $A(t)$.
Substituting (\ref{Josphase-general}) in (\ref{GenEq}) and carrying out the averaging over the rapid oscillations of frequency $v$ we obtain
$$
|A|^2=\frac{1}{4}\int \int dt_1 dt_2 G(t-t_1)G^\ast (t-t_2)\{J_0[|A|(t_1)]J_0[|A|(t_2)]
$$
\begin{equation} \label{A-dependence}
+J_2[|A|(t_1)]J_2[|A|(t_2)]\}\cos[\psi(t_1)-\psi(t_2)],
\end{equation}
where $J_n(x)$ are the Bessel functions, and the kernel $G(x)$ is the Green function of the following homogeneous equation
\begin{equation} \label{Green function}
\ddot G(t)+(2iv+\alpha)\dot G(t)-(v^2-i\alpha v)G(t)=0.
\end{equation}

Similarly we calculate the dc current $j$ flowing in the system
$$
j=\overline{\sin[A(t)e^{ivt}+B(t)e^{-ivt}+vt+\psi(t)]}=
$$
\begin{equation} \label{Current}
2\int_0^{\infty} dtIm G(t)\frac{J_1[|A|(t)]J_0[|A|(t)]}{|A|(t)}\cos[\psi(t)-\psi(0)]
\end{equation}

Thus, one can see that all electrodynamic properties of small Josephson junctions in the phase diffusion regime are determined by the specific correlation function, i.e. $\rho(t)~=~<\cos(\psi(t)-\psi(0))>$.

\section{III. Time-dependent correlation function of the Josephson phase}

In order to obtain the time-dependent correlation function of the Josephson phase we write $\psi(t)$ as
\begin{equation} \label{Psi}
\psi(t)=\int dx R(t-x)\xi(x),
\end{equation}
where the kernel $R(t)$ is the Green function of the following homogeneous equation
\begin{equation} \label{R-Green}
\ddot R(t)+\alpha \dot R(t)=0.
\end{equation}
By making use of the method proposed and elaborated in \cite{IngNaz,Fistul} we obtain the correlation function $\rho(t)$ in the following form
\begin{equation} \label{Correlationfunction}
\rho(t)=\exp \left \{-\int \frac{d\tau}{\tau_0}\int d \xi F(\xi) \left\{1-e^{i\tau_0\xi[R(t-\tau)-R(-\tau)]} \right\} \right\}~,
\end{equation}
where $F(\xi)$ and $\tau_0$ are the distribution function and the correlation time of the current noise, accordingly. Since the $R(t)$ is presented as $R(t)=\frac{(1-e^{-\alpha t})}{\alpha} \theta(t)$ we obtain in the limit of large dissipation
\begin{equation} \label{Correlationfunction-1}
\rho(t)=\exp (-\delta |t|)~, \delta \ll \alpha~,
\end{equation}
where the parameter $\delta=\frac{\tau_0}{2\alpha^2}\int d\xi \xi^2 F(\xi)$ determines the decay of the Josephson phases correlation function. In the opposite regime of $\delta \gg \alpha$ we obtain
\begin{equation} \label{Correlationfunction-2}
\rho(t)=\exp (-\delta \alpha^2 |t|^3/3)~, \delta \gg \alpha~.
\end{equation}
\section{IV. The $I$-$V$ curves of small Josephson junctions: phase-diffusion regime}

First, we notice that the function $|A(t)|$ smoothly depends on time in respect to both the kernel $G(t)$ and the correlation function $\rho(t)$. Moreover, the kernel $G(t)$ has a simple form: $G(t)=\frac{1}{\alpha}e^{ivt}$. By making use of this assumption and taking into account an explicit expression for $G(t)$  we rewrite the Eqs. (\ref{A-dependence}) and (\ref{Current}) as

\begin{equation} \label{A-dependence-1}
|A|^2=\frac{J_0^2[|A|]+J_2^2[|A|]}{4\alpha^2}\int_{-\infty}^{\infty} dt e^{ivt}\cos[\psi(t)-\psi(0)]
\end{equation}
and
\begin{equation} \label{Current-2}
j=\frac{2J_1[|A|]J_0[|A|]}{\alpha|A|}\int_0^{\infty} dt \sin(vt)\cos[\psi(t)-\psi(0)]
\end{equation}
Since we are interested in averaged quantity only, the dc $I$-$V$ curve can be expressed through the voltage dependent correlation time $\tau(v)=\tau_1(v)+i\tau_2(v)$, where
\begin{equation} \label{CorrelationTime-1}
\tau_1(v)=\left <\int_{0}^{\infty} dt \cos(vt)\cos[\psi(t)-\psi(0)]\right>
\end{equation}
and
\begin{equation} \label{CorrelationTime-2}
\tau_2(v)=\left <\int_{0}^{\infty} dt \sin(vt)\cos[\psi(t)-\psi(0)]\right>
\end{equation}
By making use of Eqs. (\ref{Correlationfunction-1}) and (\ref{Correlationfunction-2}) we obtain that $\tau_1$ approaches to the finite value for small values of voltage $v$ as
\begin{equation}
\tau_1(0)=
\begin{cases}
\frac{1}{\delta}
& \text{if }
\delta
\ll \alpha
\\
\Gamma(1/3)(9\delta \alpha^2)^{-1/3}
& \text{if }
\delta
\gg \alpha
\end{cases} \label{tau1}
\end{equation}
In the opposite limit of large values of $v$ the $\tau_1$ decreases as $1/v^2$ for overdamped junctions ($\delta \ll \alpha$) and it becomes exponentially small for underdamped junctions ($\delta \gg \alpha$). The correlation time $\tau_2$ linearly increases for small values of $v$, and decreases as $1/|v|$  for large values of voltage $v$.

Next we analyze the Eqs. (\ref{A-dependence-1}) and (\ref{Current-2}) determining the current-voltage characteristics of a small Josephson junction.
In the limit of a small value of $\tau_1(v)$ or more precisely $\tau_1 \ll \alpha^2$ the amplitude of Josephson phase oscillations $|A|$ is small, and expanding the Bessel functions over a small argument $A$ we obtain
\begin{equation} \label{I-V curves-small}
j(v)=\frac{\tau_2(v)}{\alpha}, ~\tau_1(v) \ll \alpha^2
\end{equation}
In this regime the current $I$ increases linearly in the region of small voltages, and in the limit $\delta \ll \alpha$ we recover Eq. (\ref{IVcurve-largedissipation}).
In the opposite regime, $\tau_1(v) \gg \alpha^2$, the amplitude of Josephson phase oscillations becomes large but the current $I$ is still strongly suppressed by oscillations of Bessel functions. In this strongly nonlinear regime the averaged value of $I$ is expressed through the parameters $\tau_1$ and $\tau_2$ as
\begin{equation} \label{I-V curves-large}
j(v)=\tau_2(v)\left (\frac{\alpha}{\tau_1^2} \right )^{1/3} \exp \left [-\frac{1}{4} \left(\frac{\tau_1}{4\alpha^2} \right)^{2/3} \right], ~\tau_1 \gg \alpha^2
\end{equation}
Thus, one can see that in the low dissipative junctions ($\alpha<<1$) the linear resistance is strongly (exponentially) suppressed.

\section{V. Discussion and Conclusions}
A theoretical study of the low-voltage resistive state of small Josephson junctions has been developed. In such junctions as $E_J \ll k_B T$ the thermal current fluctuations induce a stochastic time dependence of the Josephson phase. These fluctuations of the Josephson phase destroy the superconducting state and a specific resistive state occurs. The $I$-$V$ curve in such so-called Josephson phase diffusion regime crucially depends on the dimensionless dissipation parameter $\alpha$. As this parameter is large, i.e. $\alpha \gg 1$, the Josephson phase dynamics is strongly overdamped and the amplitudes of self-excited Josephson phase oscillations are small, therefore one can apply perturbation analysis, and the $I$-$V$ curve is described by Eq. (\ref{IVcurve-largedissipation}) or  Eq. (\ref{I-V curves-small}). Moreover, in the case of a large dissipation $V_m=\delta V_p$ and $I_m=I_c/(\alpha \delta)$ (see Fig. 1). Notice here, that introduced parameters $\delta$ and $\alpha$ can be expressed through the physical characteristics of a junction as $\delta=(2\pi k_B T) R_n/(V_p \Phi_0)$ and $\alpha=V_p/(I_cR_n)$, where $R_n$ is the quasi-particle resistance, $I_c$ is the nominal critical current, and $\Phi_0$ is the magnetic flux quantum \cite{FUK}.

In an opposite regime of a strongly underdamped junction, i.e. $\alpha \ll 1$, the amplitudes of self-excited Josephson oscillations are large, and the perturbation analysis can not be applied. Instead of that I used \emph{the method of averaging} elaborated previously in Ref. \cite{Gabriele,FL}. In this a strongly non-linear regime the $I$-$V$ curve is described by Eqs. (\ref{I-V curves-large}) and (\ref{tau1}). In this underdamped regime the both values $V_m=V_p \delta (\alpha/\delta)^{2/3}$ and $I_m \simeq I_c \exp{[-4^{-5/3}(\delta \alpha^8)^{-2/9})]}$ are strongly suppressed in respect to the overdamped case.

Finally, we notice that the crossover between these two  regimes is determined by the ratio of the parameters $\delta$ and $\alpha$. Since this ratio depends on the critical current value $I_c$, such a crossover can be observed experimentally in a single setup just by application and variation of an external magnetic field.


\begin{references}

\bibitem{Gabriele} M. V. Fistul and G. F. Giuliani, Phys. Rev. B \textbf{56}, 788 (1997).

\bibitem{Tinkham} M. Tinkham, Introduction to Superconductivity (McGraw-Hill, New York, 1996), 2nd ed.

\bibitem{Haviland} E. Chow, P. Delsing, and D. B. Haviland,
Phys. Rev. Lett. \textbf{81}, 204 (1998).

\bibitem{Aver-Likh} D. V. Averin, A. B. Zorin, and K. K. Likharev, Sov. Phys.
JETP \textbf{61}, 407 (1985); D. V. Averin and K. K. Likharev, J.
Low Temp. Phys. \textbf{62}, 345 (1986) .

\bibitem{Matveev} K. A. Matveev, M. Gisself\"alt, L. I. Glazman, M. Jonson, and R. I.
Shekhter, Phys. Rev. Lett. \textbf{70}, 2940 (1993).


\bibitem{AH} V. Ambegaokar and B. I. Halperin,
Phys. Rev. Lett. {\bf 22}, 1364 (1969).

\bibitem{Mooij} J. E. Mooij, and C. J. P. M. Harmans, N. J. Phys. \textbf{7}, 219 (2005)

\bibitem{Astafiev} O. V. Astafiev,	L. B. Ioffe,	S. Kafanov,	Yu. A. Pashkin,	K. Yu. Arutyunov,	 D. Shahar,	O. Cohen	and  J. S. Tsai,
    Nature \textbf{484},355 (2012).

\bibitem{MartKautz} J. M. Martinis and R. L. Kautz, Phys. Rev. Lett.
\textbf{63}, 1507 (1989); R. L. Kautz and J. M. Martinis, Phys.
Rev. B \textbf{42}, 9903 (1990).

\bibitem{FUK} Y. Koval, M. V. Fistul and A. V. Ustinov, Phys. Rev. Lett. \textbf{93}, 087004 (2004).

\bibitem{Vion-96} D. Vion, M. G\"otz, P. Joyez, D. Esteve, and M. H. Devoret,
Phys. Rev. Lett. {\bf 77} 3435 (1996).

\bibitem{IngNaz} G.-L. Ingold and Yu. V. Nazarov, in \emph{Single Charge Tunneling},
ed. H. Grabert and M. H. Devoret, NATO ASI, Ser. B, Vol. 294
(Plenum, New York, 1991).

\bibitem{GrIngPaul} H. Grabert, G.-L. Ingold and B. Paul, Europhys. Lett.
{\bf 44}, 360 (1998).

\bibitem{FL} M. P. Lisitskiy and M. V. Fistul, Phys. Rev. B \textbf{81}, 184505 (2010).

\bibitem{Fistul} M. V. Fistul, Sov. Phys. JETP \textbf{69}, 209 (1989).

\end{references}
\end{document}